\documentclass[aps,prb,reprint,superscriptaddress]{revtex4-2}
\usepackage{amsmath}
\usepackage{amssymb}
\usepackage{graphicx}
\usepackage{threeparttable}
\usepackage{xcolor}
\begin{document}

\title{Two-orbital spin-fermion model study of ferromagnetism in honeycomb lattice}
\author{Kaidi Xu}
\author{Di Hu}
\author{Jun Chen}
\author{Haoshen Ye}
\author{Lin Han}
\author{Shan-Shan Wang}
\email{wangss@seu.edu.cn}
\author{Shuai Dong}
\email{sdong@seu.edu.cn}
\affiliation{Key Laboratory of Quantum Materials and Devices of Ministry of Education, School of Physics, Southeast University, Nanjing 211189, China}
\date{\today}

\begin{abstract}
The spin-fermion model was previously successful to describe the complex phase diagrams of colossal magnetoresistive manganites and iron-based superconductors. In recent years, two-dimensional magnets have rapidly raised up as a new attractive branch of quantum materials, which are theoretically described based on classical spin models in most studies. Alternatively, here the two-orbital spin-fermion model is established as a uniform scenario to describe the ferromagnetism in a two-dimensional honeycomb lattice. This model connects the magnetic interactions with the electronic structures. Then the continuous tuning of magnetism in these honeycomb lattices can be predicted, based on a general phase diagram. The electron/hole doping, from the empty $e_{\rm g}$ to half-filled $e_{\rm g}$ limit, is studied as a benchmark. Our Monte Carlo result finds that the ferromagnetic $T_{\rm C}$ reaches the maximum at the quarter-filled case. In other regions, the linear relationship between $T_{\rm C}$ and doping concentration provides a theoretical guideline for the experimental modulations of two-dimensional ferromagnetism tuned by ionic liquid or electrical gating.
\end{abstract}
\maketitle

\section{Introduction}
Spin-fermion model describes a scenario of itinerant electrons interacting with the local magnetic moment \cite{PhysRev.82.403, PhysRev.100.675, PhysRev.118.141}, which was first proposed to study the conductivity in ferromagnetic manganites \cite{PhysRev.100.564}. It is also sometimes referred to as the Kondo model when only one orbital is involved for itinerant electrons. Moreover, it has been adopted to describe diluted magnetic semiconductors \cite{PhysRevLett.89.277202} as well as iron-based superconductors \cite{PhysRevLett.105.107004, PhysRevB.82.045125, PhysRevLett.111.047004, PhysRevLett.109.047001} with one or more orbitals considered. The most remarkable success of the spin-fermion model is for manganites, e.g. La$_{1-x}$Sr$_{x}$MnO$_3$, where it is called two-orbital double-exchange model. By tuning the doping level of $e_{\rm g}$ orbitals and thus the subtle balance between competing interactions, the phase diagram of perovskite manganites goes through a series of transitions and consequently, exotic phenomena emerge, including colossal magnetoresistivity and multiferroicity \cite{doi:10.1143/JPSJ.33.21, RevModPhys.70.1039, PhysRevLett.74.5144, dagotto2003nanoscale}.

Recently, the discovery of intrinsic two-dimensional (2D) ferromagnetism has attracted broad attention in multidisciplinary communities \cite{doi:10.1126/science.aav4450}. Many 2D van der Waals magnets with hexagonal or honeycomb geometries, such as $MX_3$-type \cite{huang_layer-dependent_2017}, $MAX_3$-type \cite{PhysRevB.91.235425, doi:10.1021/acs.nanolett.6b03052, doi:10.1021/acs.nanolett.9b05165, PhysRevB.92.035407, gong_discovery_2017, C4TC01193G}, and $M_2X_2X_6$-type transition metal trichacogenides (TMTC) \cite{gong_discovery_2017}, have been investigated. Although in these materials the magnetic transition metal ions are also caged in distorted octahedra as in three-dimensional (3D) perovskites, the adjacent octahedra are connected through an edge-sharing or face-sharing manner, instead of the common corner-sharing manner in perovskites. So far, most pioneering theoretical works on 2D magnets employed classical spin models (e.g., the Ising model or Heisenberg model with some auxiliary components), which can capture the essence of magnetic ground states. 

Using some sophisticated techniques such as ionic liquid and electrical gating,  several recent experiments demonstrated the tuning of ferromagnetism and magneto-transport in these 2D magnets \cite{deng_gate-tunable_2018, huang_electrical_2018, jiang_controlling_2018, wang_electric-field_2018}. In addition, via the electron doping from organic intercalation, the ferromagnetism of Cr$_2$Ge$_2$Te$_6$ can be significantly enhanced, i.e. its $T_{\rm C}$ is raised  to $\sim200$ K \cite{doi:10.1021/jacs.9b06929}. Besides these artificial tunings, an interesting fact is  that the $T_{\rm C}$'s of Mn$X_3$'s (predicted to be hundreds Kelvin) are one order of magnitude higher than those of neighboring Cr$X_3$'s. All these phenomena suggest that the charge/orbital degrees of freedom play a vital role in determining the magnetism in 2D magnets. However, those classical spin models themselves are incapable of tuning the charge degree of freedom and fail to capture the physics governed by the electronic structures and quantum fluctuation. Therefore, a more microscopic quantum model, e.g. the spin-fermion model, should be developed to unveil more novel properties and deeper physics of 2D magnets, as done for 3D quantum magnets \cite{PhysRevLett.80.845}.

\begin{figure}
\centering
\includegraphics[width=0.48\textwidth]{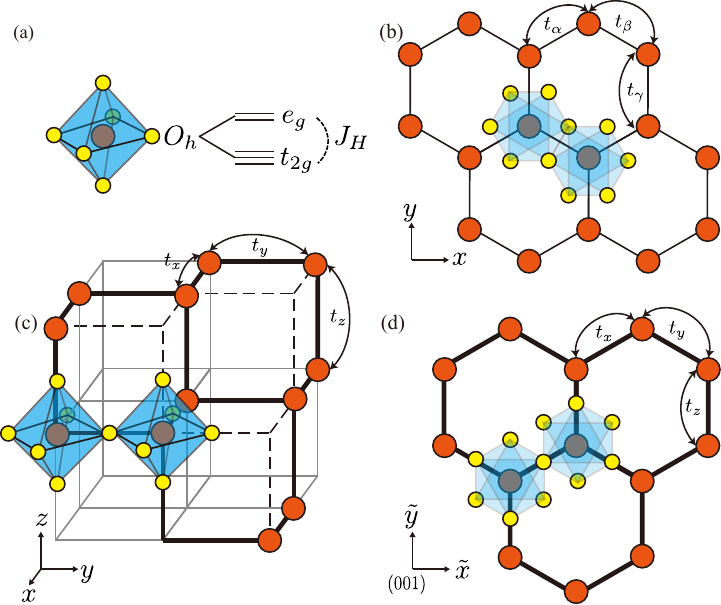}
 \caption{Comparison of low-dimensional honeycomb magnets and [111]-bilayer of $AB$O$_3$ perovskites. (a) The local octahedral structure. The red and yellow circles represent the magnetic cations and ligand anions, respectively. The $d$-orbitals split into the $e_{\rm g}$ and $t_{\rm 2g}$ orbitals in an undistorted octahedral crystal field in the absence of spin-orbit coupling (SOC). The neighboring octahedra are connected in an edge-sharing manner ($M$-$X$-$M$ bond angle $\sim90^{\circ}$), which is the common case for $MX_3$, $MAX_3$, and $M_2 A_2 X_6$ type van der Waals magnets.  (c)  The neighboring $M$O$_6$ octahedra share a common corner ($M$-O-$M$ bond angle $\sim180^{\circ}$) in perovskites. The bold solid lines denote the [111]-oriented bilayer, which forms a buckled honeycomb lattice. (d) Top view of the [111] bilayer. The hopping integrals of $e_{\rm g}$ electrons are schematically shown by double arrows.}
    \label{fig:geo}
\end{figure}

For manganites and iron-chalcogenides/pnictides, the nearly cubic/square geometries make their octahedra or tetrahedra coordinates of $e_{\rm g}$ and $t_{\rm 2g}$ wave functions uniform through the whole lattice. However, for these hexagonal or honeycomb lattices, when connecting the octahedra/double-pyramid via edge-sharing or face-sharing manner, the $e_{\rm g}$ (and others) wave functions no longer form a representation for the corresponding point group, i.e., the basis transform as $e_{\rm g}$ cannot form a representation matrix to transform hopping integral to another hopping path. Although the [111]-bilayer of perovskites can mimic the honeycomb geometry  \cite{xiao_interface_2011, PhysRevB.84.201104}, as shown in Fig.~\ref{fig:geo}(c),  the $e_{\rm g}$ basis set and hopping integrals do not change from its 3D form, since the neighboring octahedra are still conner-sharing. Therefore, those  real hexagonal or honeycomb magnetic lattices are different from the [111] bilayers.

Here, we develop a two-orbital spin-fermion model with minimal hopping parameters for the honeycomb lattice with neighboring octahedra connected in an edge-sharing manner, to describe the ferromagnetism of $MX_3$ and other similar monolayers. We confine the orbital configurations with half-filled $t_{\rm 2g}$ triplets and partially-occupied $e_{\rm g}$ doublets, which can generally describe those cases with high-spin $d^3$-$d^5$ transition metal ions.  The hopping parameters can be extracted from density functional theory (DFT) calculations, which can be reduced to two-independent values by group theory in the ideal limit.

\section{Model \& Methods}
In general, the spin-fermion Hamiltonian can be expressed as \cite{dagotto2003nanoscale}:
\begin{equation}
\begin{split}
H= &-\sum_{<ij>} t_{ij} c^{\dagger}_{i\sigma} c_{j\sigma} + J_{\rm H} \sum_{i} c^{\dagger}_{i\sigma} \sigma c_{i\sigma} \cdot S_{i}
\\
&+ J_{ex} \sum_{<ij>} S_i \cdot S_j +A_{z} \sum_{i} (S^{z}_{i})^2,
\label{de}
\end{split}
\end{equation}
where $t_{ij}$ is the spin-conserved hopping integral for itinerant $e_{\rm g}$ electron, $J_{\rm H}$ is the Hund's coupling coefficient, $J_{ex}$ is the superexchange parameter and $S$ is the classical spin vector for $t_{\rm 2g}$. $c^{\dagger}_{i\sigma} \sigma c_{i\sigma}$ represents the spin operator for the itinerant electron in the quantum level. $A_{z}$ parameterizes single-ion magnetocrystalline anisotropy.

The spin vector $S$ can be expressed as $S$($\sin\theta\cos\phi$, $\sin\theta\sin\phi$, $\cos\theta$), where $\theta$ and $\phi$ are the polar and azimuth angles, respectively. In the strong Hund's coupling limit which is generally valid for most $3d$ electronic materials with the high spin configurations (e.g., $S=3/2$ in our case) \cite{dagotto2003nanoscale}, the spin of the itinerant electron is confined to align parallel with the localized spin. Then the first two items in Eq.~(\ref{de}) can be simplified into a spinless form: $\sum_{<ij>} t_{ij}\Omega_{ij}c^{\dagger}_ic_j$, where the Berry phase $\Omega_{ij}$ from the spin texture can be expressed as:
\begin{equation}
\Omega_{ij}=\cos\frac{\theta_i}{2}\cos\frac{\theta_j}{2}+\sin\frac{\theta_i}{2}\sin\frac{\theta_j}{2}\exp^{-i(\phi_i - \phi_j)}.
\end{equation}
For a plain ferromagnetic background, $\Omega_{ij}=1$.

For those itinerant electrons involving the orbital degree of freedom, the hopping parameter $t$ can be orbital-/orientation-dependent,  which is defined as:
\begin{equation}
t^{\mu \mu'}_R=\langle \psi^{\mu}(r)|H|\psi^{\mu'}(r-R) \rangle,
\end{equation}
where $\mu$/$\mu'$ labels the orbital basis, $\psi(r)$ is the wave function, and $R$ is the vector along the hopping path. Considering the symmetries of given lattice, the hopping coefficients are not fully independent, which are related by:
\begin{equation}
t(\hat{g}_n R)=D (\hat{g}_n) t(R) [D(\hat{g}_n)]^{\dagger},
\label{t}
\end{equation}
where $\hat{g}_n$ is a symmetry operation of given point group.

In perovskite manganites involving three hopping direction ($x$/$y$/$z$) and two $e_{\rm g}$ orbitals $\{x^2-y^2$, $3z^2-r^2\}$, in principle, there are nine components of $t$. Benefiting from the $O_{h}$ point group of cubic lattice, the independent components are reduced to only one $t_0$ by the relationship (Eq.~\ref{t}), while all others are proportional to $t_0$. Although in most manganites the lattices are not ideally cubic but distorted to an orthorhombic, rhombohedral, or tetragonal pseudocubic one, the cubic approximation remains a successful choice. 

For those 2D magnets, the on-site crystal field remains (distorted) octahedral, leading to the same splitting between low-lying $t_{\rm 2g}$ and higher $e_{\rm g}$ orbitals.  However, the key difference for these honeycomb lattices is that their neighboring octahedra are connected in an edge-sharing manner.  Thus, the local orbital basis set $\{x^2-y^2$, $3z^2-r^2\}$ no longer form a global representation for the honeycomb lattice, whose point group belongs to $D_{3d}$. In the honeycomb lattice, the hopping path can be related by three-fold rotation symmetry (i.e., $R_{\beta(\gamma)} = C^{1(2)}_3 R_{\alpha}$). However, accompanying the $C_3$ transformation, the $e_{\rm g}$ orbital base will inevitably mix with the $t_{\rm 2g}$ $xy$ component. In fact, the  full in-plane rotation matrix of $d$-orbitals can be written as $R=R_{{xy, x^2-y^2}}\oplus R_{{3z^2-r^2}} \oplus R_{xz,yz}$, as explained in Supplementary Material (SM) \cite{sm}. Therefore, the hopping $t_R$ does not obey the relationship as in the cubic case anymore, which needs a careful reestablishment. 

According to the Slater-Koster method \cite{PhysRev.94.1498}, the hopping integral contributed via the $d-p-d$ orbital hybridization can be approximated by the second-order perturbation $t_{dpd} = \frac{t_{pd}^{\prime}t_{pd}}{\Delta_{pd}}$, where $t_{pd}$ is the hopping coefficient and $\Delta_{pd}$ is the energy difference between the $p$ and $d$ orbitals \cite{dpd}. Here, after the orbital hybridization, the $e_g$ and ligand-$p$ orbitals form a doublet composed by $|a\rangle$ ($x^2-y^2$ and $p$ character) and $|b\rangle$ ($3z^2-r^2$ and $p$ character). The new states should respect the same symmetry with the $e_g$ orbitals \cite{khomskii_2014}. Then the nearest-neighboring $d-d$ hopping integral obtained by Slater-Koster coefficients in the honeycomb lattice with edge-sharing octahedra resembles the following form  \cite{PhysRev.94.1498}:
\begin{equation}
\begin{aligned}
    t_{\alpha} &=
        \begin{bmatrix}
            t^{aa 2} & 0\\
            0 & t^{bb 2} 
        \end{bmatrix},\\
    t_{\beta} &=
        \begin{bmatrix}
            t^{aa 1} & t^{ab}\\
            t^{ab} & t^{bb 1}
        \end{bmatrix},\\
    t_{\gamma} &=
        \begin{bmatrix}
            t^{aa 1} & -t^{ab}\\
            -t^{ab} & t^{bb 1}
        \end{bmatrix},
    \end{aligned}
\end{equation}
where $\alpha$/$\beta$/$\gamma$ denote the three hopping paths as indicated in Fig.~\ref{fig:geo}(b). Then in the momentum space, a four-band Hamiltonian can be obtained and solved analytically, as shown in SM \cite{sm}.

For a $D_{3d}$ point group with site symmetry being $C_{3v}$ and basis set spanned by a general doublet $\{|a\rangle, |b\rangle \}$, the following symmetric arguments can be proved \cite{xu_orbital-active_2022}: 

First, at the $\Gamma$ point, the eigenvectors transform as two 2D irreducible representations, namely,  $a_{\rm 1g} \otimes e_{\rm g} = e_{\rm g}$ and  $a_{2u} \otimes e_{\rm g} = e_{\rm u}$. Therefore, this symmetry enforces two quadratic touching pairs of eigenvalues at the $\Gamma$ point.

Second, at the $K$ (and $K'$) point, the eigenvector can be decomposed into $\bigoplus K(K') = a_1 \oplus a_1 \oplus e$. Therefore, there is a band degenerate pair at the $K$ ($K'$) point with opposite chirality (\textit{i.e.}, a Weyl point).

With these two symmetry constraints, one could obtain some relationships between the hopping coefficients by comparing the eigenvalues of $H$ of ferromagnetic state at the $\Gamma$ and $K$ points. For the degeneracy at the $\Gamma$ point, one has:
 \begin{equation}
2t^{bb1} + t^{bb2}=2t^{aa1}+t^{aa2}.
\label{g}
\end{equation}
For the degeneracy at the $K$ point, one arrives:
\begin{equation}
\begin{split}
t^{aa2} &-t^{aa1}=t^{bb1}-t^{bb2},\\
\sqrt{3}t^{ab}&=\sqrt{(t^{aa2}-t^{aa1})(t^{bb1}-t^{bb2})}\\
&=|t^{bb1}-t^{bb2}|=|t^{aa2}-t^{aa1}|.
\end{split}
\label{k}
\end{equation}

Besides the analytic solution for the ferromagnetic background, the Hamiltonian can also be numerically solved using an unbiased Monte Carlo simulation. The details of Monte Carlo algorithm for such a spin-fermion model can be found in SM \cite{sm}, which has been proved to be successful to simulate the complex magnetism in manganites and others \cite{dagotto2003nanoscale}.

\section{Results}
The hopping coefficients in real materials are contributed by both the direct $d-d$ interaction and the effective interaction mediated by ligand-$p$ orbitals \cite{PhysRevB.88.085433}, while aforementioned symmetric relationships should work for both cases. To verify this point, the hopping coefficients of a protype 2D honeycomb magnet MnF$_3$ are calculated by the maximally localized Wannier functions (MLWFs) using the Wannier90 code \cite{mostofi_wannier90_2008}.  For comparison, more hopping coefficients of other 2D honeycomb materials taken from literature are also listed in Table~\ref{tab}. All of them fit Eqs.~\ref{g} and \ref{k} very well. Thus, the independent hopping coefficients between $e_{\rm g}$ orbitals in the 2D honeycomb lattice can be reduced to two parameters, e.g. $t^{bb1}$ and $t^{bb2}$, different from the perovskite cases where the hopping integrals can be determined by a single parameter.

\begin{table}
\caption{The nearest-neighbor transfer integral between $e_{\rm g}$ orbitals fitted by MLWFs (in units of meV). }
\begin{tabular}{l|ccccc}
\hline
 & MnF$_3$ & AuF$_3$\cite{PhysRevB.99.041101} & AuCl$_3$\cite{PhysRevB.99.041101} & AuI$_3$\cite{PhysRevB.99.041101} & PdPS$_3$ \cite{PhysRevB.97.035125}\\ \hline
$t^{bb1}$ & 14.5 & 224 & 60  & 44 & 82\\ \hline
$t^{bb2}$ & 227  & -50 & -16 & 53 & 70\\ \hline
$t^{aa1}$ & 156  & 41  & 9   & 50 & 72\\ \hline
$t^{aa2}$ & -56   & 313 & 85  & 41 & 87\\ \hline
$t^{ab}$  & 122  & 158 & 44  & 5  & 9\\ \hline
\end{tabular}
\label{tab}
\end{table}

\begin{figure}
\centering
\includegraphics[width=0.48\textwidth]{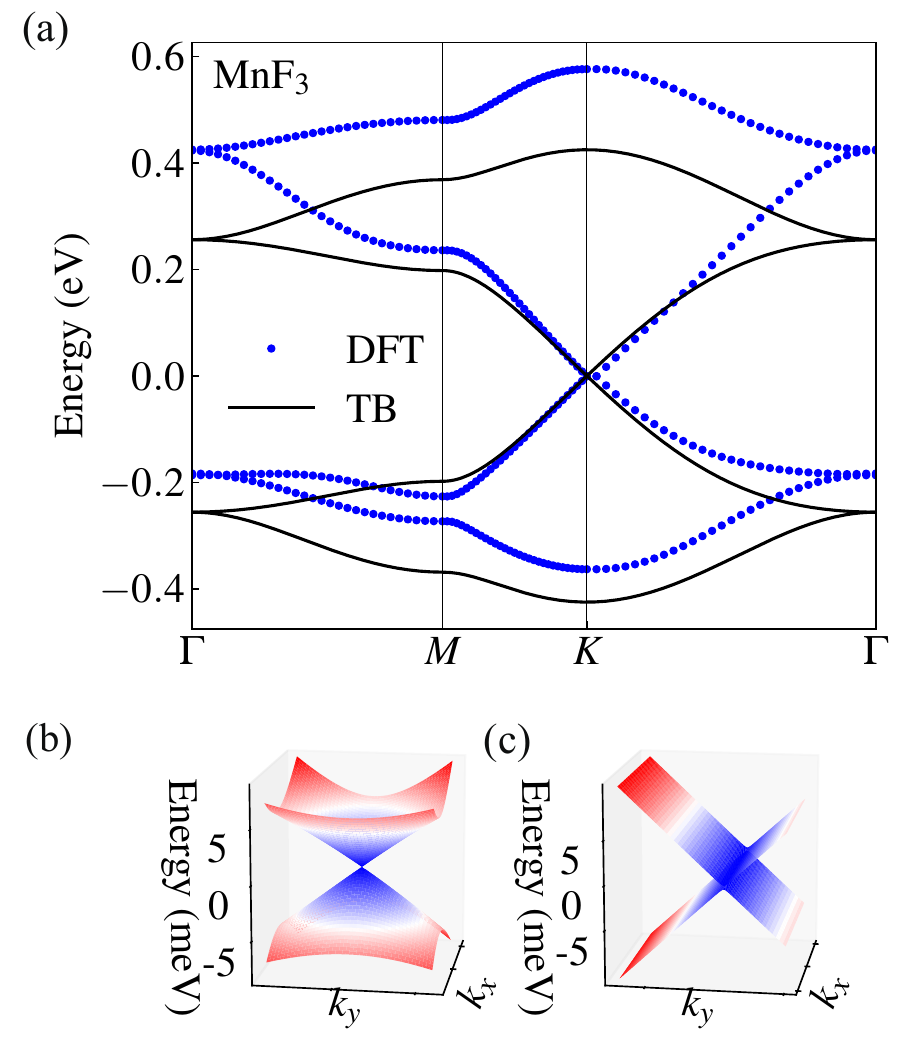}
\caption{Electronic band structures of two-orbital model on the honeycomb lattice. Here the plain ferromagnetic background is used. (a) A comparison between DFT band structure and model band structure for MnF$_3$ monolayer.  Here the values of $t^{bb1}$ and $t^{bb2}$ are listed in Table~\ref{tab}, obtained from the MLWFs fitting. In DFT calculation, the electron-hole symmetry is broken, leading to wider conducting bands than valence bands. (b) The 3D dispersion curve near a single Weyl point at $K$ for MnF$_3$. (c) The 3D dispersion curve for a Weyl nodal line when using $t^{bb1}=139$ meV and $t^{bb2}=227$ meV. }
\label{fig:topo}
\end{figure}

By tuning the values of $t^{bb1}$ and $t^{bb2}$, it can be found that $t^{bb2}$ mainly changes the energy scale and $t^{bb1}$ mainly alters the shape of band structures, as shown in Figs.~S2 and S3 in SM \cite{sm}. A typical band structure for one $e_{\rm g}$ occupancy (MnF$_3$) are shown in Fig.~\ref{fig:topo}(a). The band degeneracy at $\Gamma$ and $K$ can be clearly evidenced in both the model bands and the DFT bands, as protected by corresponding symmetries. The similarity between the model and DFT bands implies the main physics has been well captured in our model. The quantitative difference is mainly due to the breaking of electron-hole symmetry in the DFT calculations, which comes from the hoppings beyond the nearest-neighbors as well as the interaction between electrons. By tuning the values of $t^{bb1}$ and $t^{bb2}$, interestingly, both the Weyl cone and Weyl nodal line are possible near the Fermi level, as compared in Fig.~\ref{fig:topo}(b-c).

\begin{figure}
\centering
\includegraphics[width=0.48\textwidth]{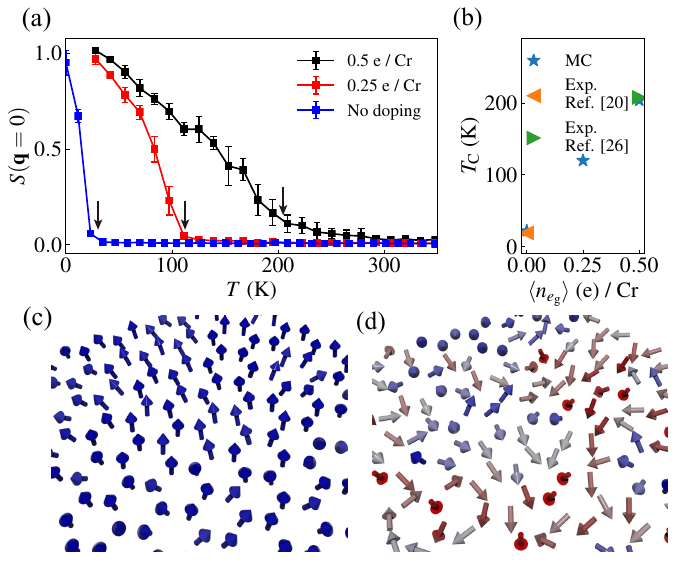}
\caption{Model simulated ferromagnetic transitions in undoped/doped CrGeTe$_3$ monolayers. The static structure factor of $t_{\rm 2g}$ spins is defined as $S(q)=\sum_{ij}\langle S_i \cdot S_j\rangle\exp[-i q \cdot (R_i - R_j)]$, where $\langle S_i \cdot S_j \rangle$ is the spin correlation function in real space. (a) The $S(q=0)$ curves for different doping levels as a function of temperature. The $T_{\rm C}$ can be significantly enhanced upon electron doping. (b) The Monte Carlo $T_{\rm C}$ as a function of doping level, in comparison with the values obtained in the organic intercalation experiments. (c-d) The Monte Carlo snapshots at (c) $42$ K and (d) $300$ K, respectively, for the $0.5$ electron/Cr doping case.}
\label{fig:tj}
\end{figure}

In the following, the ferromagnetic transition will be studied using the two-orbital spin-fermion model on the honeycomb lattice. For better comparison with available experimental data, here we choose CrGeTe$_3$ monolayer as the starting material, which owns the $S=3/2$ high-spin configuration, as supported by the magnetic moment $2.75$ $\mu_{\rm B}$/Cr (slightly lower than the ideal $3$ $\mu_{\rm B}$ due to near 90$^{\circ}$ $dp\sigma$ hybridization \cite{doi:10.1021/jacs.9b06929}). Experimentally, CrGeTe$_3$ monolayer is ferromagnetic with a low Curie temperature $T_{\rm C}=21$ K \cite{gong_discovery_2017}. This fact implies a ferromagnetic superexchange between $t_{\rm 2g}$ spins, i.e., $J_{ex}=-2.71$ meV according to Ref.~\cite{gong_discovery_2017}. With this $J_{ex}$ and single-axis magnetocrystalline anisotropy ($A_z=-0.5$ meV with $z$ being the easy-axis from our DFT calculation), our Monte Carlo simulation leads to a $T_{\rm C}\sim25$ K, as shown in Fig.~\ref{fig:tj}(a).

Then the electron doping is considered. Using the MLWFs fitting, the hoping intensity $t^{bb1}$ and  $t^{bb2}$  for CrGeTe$_3$ monolayer are obtained as $16.2$ meV and $215$ meV, respectively.  In addition, according to our DFT calculation,  the shape of $e_{\rm g}$ bands is almost unchanged after doping, consistent with the experiment evidences \cite{wang_electric-field_2018, doi:10.1021/jacs.9b06929, verzhbitskiy_controlling_2020}. Thus, the $t^{bb1}$ and  $t^{bb2}$ coefficients will be fixed in our following model simulation. After doping one electron per unit cell (i.e., $0.5$ electron/Cr, which corresponds to $2.46\times10^{14}$  electron/cm$^2$ in experiments),  our MC simulation obtains a much enhanced $T_{\rm C}=204$ K, which is compatible with the organic intercalation experiments at this doping level with $T_{\rm C} =208$ K \cite{doi:10.1021/jacs.9b06929}. Noting that in the experiment, after the organic intercalation CrGeTe$_3$ layers are nearly isolated to each other, which can mimic the monolayer case.

\begin{figure}
\centering
\includegraphics[width=0.48\textwidth]{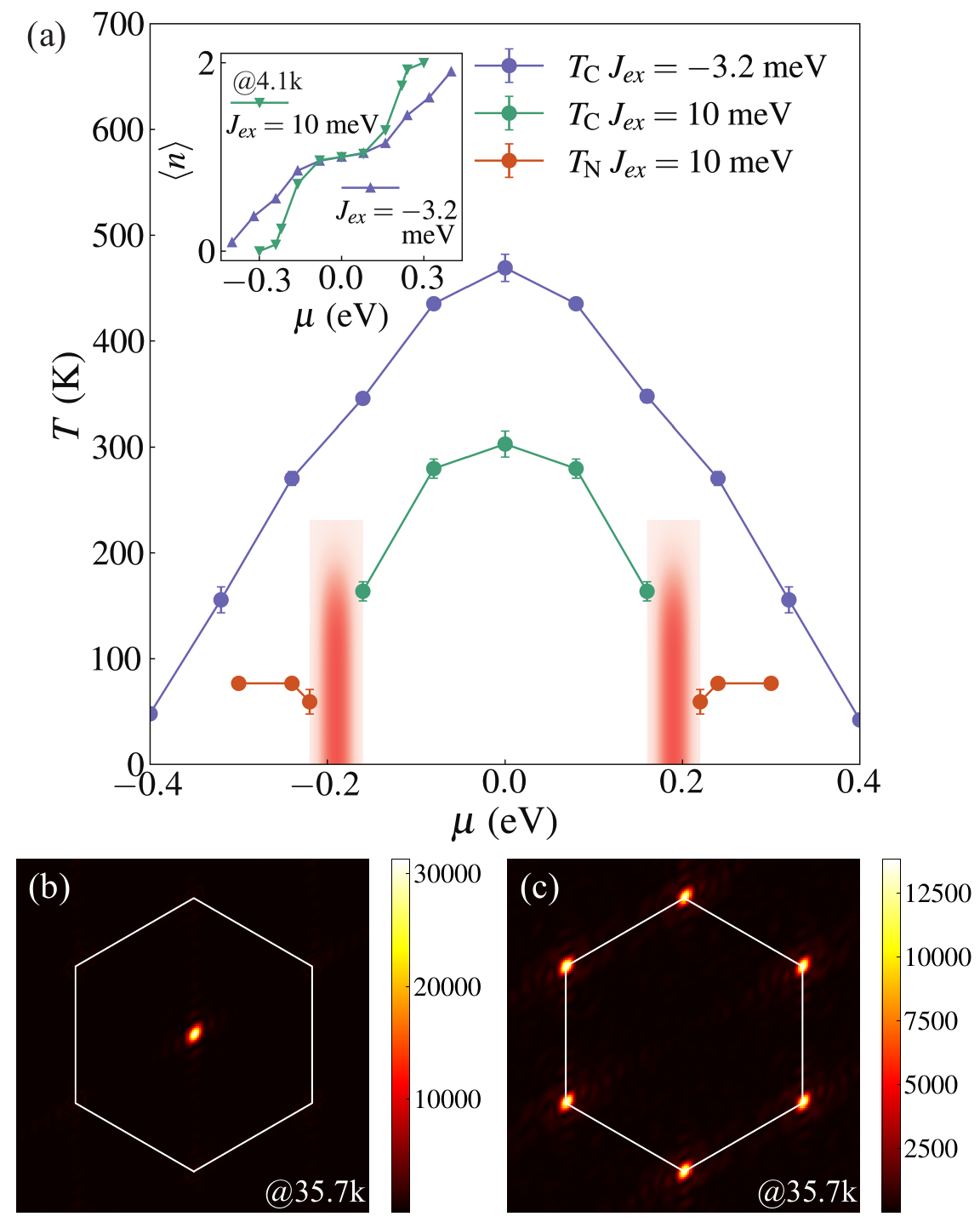}
\caption{(a) The evolution of magnetic transition temperatures ($T_{\rm C}$ or $T_{\rm N}$) as a function of chemical potential $\mu$. Here two types of $J_{ex}$ are considered: $J_{ex}=-3.2$ meV and $J_{ex}=10$ meV.  The former leads to paramagnetic-ferromagnetic transitions in the whole range, while the latter can lead to paramagnetic-antiferromagnetic transitions in two ends.  Inset: The corresponding $e_{\rm g}$ electron density ($n$) as a function of  $\mu$ at $4.1$ K. (b-c) Typical spin structure factors at $36.7$ K. (b) The ferromagnetic state with Bragg peak centered at the $\Gamma$ point and (c) the Néel antiferromagnetic state with Bragg peaks resided on the corner of 1st Brilliouin Zone.}
\label{fig:Tc}
\end{figure}

Above results have demonstrated the successful description of ferromagnetic transitions in doped CrGeTe$_3$ monolayers, which allows us to go further to more honeycomb magnets, especially those not well studied yet. In the following, we will demonstrate the continuous tuning of ferromagnetism in the $MX_3$ system. We start with the $3d^4$ configuration (i.e.,  Mn$X_3$). The transfer integral coefficients ($t^{bb1}$ and  $t^{bb2}$) are chosen to fit the \textit{ab initio} data of MnF$_3$. Empirically, F$^-$ anion is expected to generate larger crystal field and smaller SOC comparing with other halogen Br$^-$, Cl$^-$, and I$^-$ \cite{khomskii_2014}, which make MnF$_3$ an ideal candidate as a benchmark. In the full hole-doped case, the system becomes $3d^3$, which can mimic Cr$X_3$. Considering the $T_{\rm C}$ of CrF$_3$ determined by numerical studies \cite{C5TC02840J}, $J_{ex}$ is chosen as $-3.2$ meV and easy-axis type magnetocrytalline anisotropy energy $A_{z}=-0.5$ meV, which leads to $T_{\rm C}=35$ K. Then by tuning the chemical potential continuously, a MC phase diagram for $MX_3$ from $3d^3$ to $3d^5$ can be obtained, as presented in Fig.~\ref{fig:Tc}.  The value of $T_{\rm C}$ can be drastically enhanced from $35$ K (for $3d^3$) to $450$ K (for $3d^4$), in a nearly linear manner upon increasing $3d$ electron concentration. Due to the aforementioned particle-hole symmetry in our model, the $T_{\rm C}$ of $3d^{4+\delta}$ case equals to the one of $3d^{4-\delta}$, resulting in a nearly linear dropping of $T_{\rm C}$ upon electron concentration from $3d^4$ to $3d^5$ (e.g., Fe$X_3$). 

Our model simulation can also deal with the case with antiferromagnetic $J_{ex}$, which can compete with the ferromagnetic $e_{\rm g}$ hoppings (i.e., double-exchange). Also shown in Fig.~\ref{fig:Tc}, the ferromagnetism appears in the middle region ($\sim3d^4$ region), while the N\'eel-type antiferromagnetic phase appears in two ends ($\sim3d^3$ and $\sim3d^5$), also in the symmetric manner. The difference of ferromagnetism and antiferromagnetism can be clearly visualized by its order parameter ($S(q)$), as compared in Fig.~\ref{fig:Tc}(b-c).

The particle-hole symmetry of electronic bands would be slightly broken when the Hamiltonian includes higher-order hopping terms as well as Coulombic interaction, as visualized in Fig.~\ref{fig:topo}(a). As a consequence, the ferromagnetism in Fe$X_3$ ($3d^5$) is expected to be even weaker than Cr$X_3$ ($3d^3$) \cite{tomar_intrinsic_2019}. In fact, several \textit{ab-initio} studies showed controversial results on whether the ground states of Fe$X_3$ is ferromagnetic or even antiferromagnetic \cite{tomar_intrinsic_2019, doi:10.1021/acs.jpcc.1c03915}. Another possible reason to break the particle-hole symmetry in real $MX_3$ is the non-negligible changes of coefficients ($t^{bb1}$,  $t^{bb2}$, $J_{ex}$) as well as the crystalline field splitting when $M$ changes from Cr to Fe. For example, the lattice constants of CrF$_3$ and MnF$_3$ are differ by $0.22$ \AA{} \cite{doi:10.1021/acs.jpcc.2c06733, PhysRevB.97.094408}. Even so, our MC simulation of spin-fermion model can still capture the main physics and lead to a semiquantitative description of $MX_3$ monolayers. More precise description of concrete materials and subtle physics can be obtained by using more precise hopping coefficients and $J_{ex}$ for particular materials. 

Generally, above phase diagram and underlying physical mechanism are similar for both the corner-sharing and edge-sharing cases, namely the ferromagnetic $T_{\rm C}$'s are mostly related to the bandwidth and occupation level of itinerant electrons \cite{PhysRevLett.80.845, PhysRevLett.103.107204}. However, one can not straightforwardly compare their $T_{\rm C}$'s from the model aspect, since different coefficients ($t^{bb1}$/$t^{bb2}$ {\it vs} $t_0$) and coordinations (three neighbors  {\it vs} four/six neighbors) are involved.  Even so, it is unambiguous that the edge-sharing case is more favor of rich band topology inherited from its honeycomb geometry.

Our spin-fermion model is just a starting point to investigate magnetism in these 2D honeycomb magnets. Other interactions can be further implemented to describe more physical phenomena. For example, in order to describe those $4d$/$5d$ electrons with medium or low spin configurations, one may need to consider the moderate Hund's coupling and non-negligible SOC. In addition, higher order hopping terms may also be needed, considering the more spatially extended $4d$/$5d$ electron clouds. Then more exotic effects, like the Kitaev physics, may be explored with the revised spin-fermion model.

\section{Conclusion}
In summary, we proposed a general two-orbital spin-fermion model on the honeycomb lattice with edge-sharing octahedra. The relationship between orbital-orientation-dependent hopping coefficients were derived from symmetry analysis, leading to two independent parameters. Based on this model, Monte Carlo simulation was employed to investigate the evolution of magnetism upon doping, which can well reproduce the experimental gating-tunable ferromagnetism in some 2D materials. As a general tight-binding model to describe the magnetism in honeycomb lattice, our model goes beyond the classic spin model and captures deeper physics originating from electronic structures. As a starting point, our model is also adaptable to include more realistic terms to cover more interesting physics in these two-dimensional quantum materials, such as phase competition and topological bands, which will stimulate more following works in near future.

\begin{acknowledgments}
K. X would like to thank Dr. Ning Ding, Prof. Xiaoyan Yao of Southeast University, and Prof. Yukitoshi Motome of the University of Tokyo for fruitful discussions. K. X. also would like to acknowledge the kind hospitality by CNR-SPIN c/o Department of Chemical and Physical Science of University of L’Aquila (Italy) during the preparation of this manuscripts. This work  was supported by Natural Science Foundation of China (Grant Nos. 11834002 and 12104089), Natural Science Foundation of Jiangsu Province (Grant No. BK20200345), and the Big Data Computing Center of Southeast University. 
\end{acknowledgments}

\bibliographystyle{apsrev4-2}
\bibliography{cite}
\nocite{*}
\end{document}